\documentclass[aps,prd,print,superscriptaddress,amsfonts,amssymb,amsmath
,nofootinbib
]{revtex4-1}
\usepackage{ulem,color,wrapfig,graphicx}

\begin{document}

\title{Gravitational Waves in $F(R)$ Gravity: \\Scalar Waves and the Chameleon Mechanism}

\author{Taishi Katsuragawa}
\email{taishi@mail.ccnu.edu.cn}
\affiliation{Institute of Astrophysics, Central China Normal University, Wuhan 430079, China}
\author{Tomohiro Nakamura}
\email{nakamura.tomohiro@g.mbox.nagoya-u.ac.jp}
\affiliation{Department of Physics, Nagoya University, Nagoya 464-8602, Japan}
\author{Taishi Ikeda}
\email{taishi.ikeda@tecnico.ulisboa.pt}
\affiliation{CENTRA, Departamento de F\'isica, Instituto Superior T\'ecnico
– IST, Universidade de Lisboa – UL, Avenida Rovisco Pais 1, 1049 Lisboa, Portugal}
\author{Salvatore Capozziello}
\email{capozziello@na.infn.it}
\affiliation{Dipartimento di Fisica ``E. Pancini", Universit\`a di Napoli ``Federico II",
Complesso Universitario di Monte Sant’Angelo, Edificio G, Via Cinthia, I-80126, Napoli, Italy}
\affiliation{Istituto Nazionale di Fisica Nucleare (INFN) Sezione di Napoli,
Complesso Universitario di Monte Sant’Angelo, Edificio G, Via Cinthia, I-80126, Napoli, Italy}
\affiliation{Laboratory for Theoretical Cosmology,
Tomsk State University of Control Systems and Radioelectronics (TUSUR), 634050 Tomsk, Russia.}

\begin{abstract}
We discuss the scalar mode of gravitational waves emerging in the context of $F(R)$ gravity by taking into account the chameleon mechanism.
Assuming a toy model with a specific matter distribution to reproduce the environment of detection experiment by a ground-based gravitational wave observatory, 
we find that the chameleon mechanism remarkably suppresses the scalar wave in the atmosphere of Earth, compared with the tensor modes of the gravitational waves.
We also discuss the possibility to detect and constrain scalar waves by the current gravitational observatories and advocate a necessity of the future space-based observations. 
\end{abstract}


\maketitle


\section{Introduction}
\label{intro}

Gravitational waves (GWs) are now the key probes to observe high-energy astrophysical phenomena
after the success of detecting the signals from coalescing binary black holes \cite{Abbott:2016blz, Abbott:2016nmj, Abbott:2017vtc, Abbott:2017oio, Abbott:2017gyy} and binary neutron stars \cite{TheLIGOScientific:2017qsa, Monitor:2017mdv}.
Moreover,  forthcoming observational observatories are aimed to give deeper and clearer understandings of the current and early Universe \cite{Aso:2013eba, Unnikrishnan:2013qwa}.

Besides, GWs have the potential to test any gravitational theory.
Several alternative theories to general relativity have been investigated so far, and some of them predict different propagation speed and polarization modes of GWs from those of general relativity.
Thus, we need to pay more attention to the study of GWs 
as a new and formidable method to test any theory of gravity.
In principle, finding new features in GWs, 
which are unexplainable in the context of general relativity, 
we can consider those features as a sort of smoking gun of any gravitational theory beyond the general relativity.
On the other hand,
if we cannot find any deviations from the theoretical predictions of general relativity,
we can obtain constraints on modified gravity theories.
For instance, the GW speed provides us with the constraint on the several classes of modified gravities (see for example 
\cite{Lombriser:2015sxa, Lombriser:2016yzn, Ezquiaga:2017ekz, Creminelli:2017sry, Sakstein:2017xjx}).

In this context, $F(R)$ gravity is one of the modified gravity theories, and its action is replaced by the function of the Ricci scalar, instead of the Einstein-Hilbert action. Specifically, it can be considered as the straightforward extension of general relativity and several viable models of $F(R)$ gravity are consistent with observational constraints coming from GWs and other astrophysical probes  (for review, see \cite{Nojiri:2010wj, Capozziello:2011et, Annalen, Nojiri:2017ncd}).
It is widely known that $F(R)$ gravity is equivalent to a scalar-tensor theory, which includes the scalar field in addition to the metric, at the classical level \cite{Capozziello:2011et, Ntahompagaze:2017dla}.

In the models of $F(R)$ gravity for the dark energy,
the scalar field plays a role of the dynamical dark energy, instead of the cosmological constant \cite{Elmardi:2016mtp}.
For  consistency with constraints from the various observational data, 
viable $F(R)$ gravity models should satisfy the chameleon mechanism
\cite{Khoury:2003rn, Brax:2008hh, Capozziello:2007eu}.
The chameleon mechanism is one of the screening mechanisms
which suppress the so-called fifth force mediated by the scalar field.
The chameleon mechanism in $F(R)$ gravity requires that the potential of the scalar field is sensitive to the trace of the energy-momentum tensor,
which brings the environment dependence of the scalar field.
The well-designed potential predicts that the scalar field acquires the large mass and that the fifth force is suppressed in the high-density region (e.g., the Solar System).

The additional scalar field results in that the $F(R)$ gravity gives rise to  a further scalar mode of  GWs 
\cite{Capozziello:2008rq, Yang:2011cp, Rizwana:2016qdq, Gong:2017bru, Liang:2017ahj},
which is described by the fluctuation of the scalar field around the potential minimum.
Here, it is logical that 
the chameleon mechanism should also affect the perturbation of the scalar field.
In other words, we naively expect that the environment affects the scalar mode of GWs because the potential of the scalar field changes according to the medium in which the GWs propagate.
There are three possible effects on the scalar mode of GWs due to the chameleon mechanism:
at the emission \cite{Liu:2018sia, Zhang:2018prg, Jana:2018djs},
during the propagation \cite{Lindroos:2015fdt, Hagala:2016fks, Ip:2018nhl}, 
and at the detection.

In this paper, we study the effect of the chameleon mechanism on the scalar mode of GWs at the detection by ground-based GW observatories.
We derive the wave equation of the scalar field at the linear order of perturbation around the non-flat background and discuss the method to evaluate the scalar waves.
Assuming a toy model with a simplified distribution of matter field 
which mimics the detection experiments by the ground-based observatories, 
we examine the chameleon mechanism and give the analytic expressions of the amplitude of the scalar wave, comparing with the tensor modes of GWs.

Remarkably,
we will find that the scalar wave attenuates during the propagation in the atmosphere due to the chameleon mechanism and that the amplitude of the scalar wave can be considered almost zero when the scalar wave enters the GW detector.
The above results are consistent with the current observational data which do not explicitly show the extra polarization modes in the GW signals.
That is, the chameleon mechanism gives us a key understanding to explain the fact that 
no extra polarization modes have been detected so far.
We claim that the chameleon mechanism plays a vital role in the GW observations as well as the fifth-force observation and that the GW physics in $F(R)$ gravity should respect the chameleon mechanism.
We also comment on the detectability of the scalar field by the ground-based GW observatories and the necessity of  future space-based GW observatories
\cite{Kawamura:2011zz, McNamara:2008zz, Audley:2017drz, Luo:2015ght} for the direct detection of scalar waves.


\section{$F(R)$ Gravity and the Chameleon Mechanism}
\label{sec2}

\subsection{ A  scalar field coming from $F(R)$ gravity}
\label{sec2A}
We give a brief review of $F(R)$ gravity and its field equations.
We also see how the chameleon mechanism works in a particular model of $F(R)$ gravity.
The action of the generic $F(R)$ gravity model is defined as follows:
\begin{align}
S=\frac{1}{2\kappa^{2}} \int d^{4}x \sqrt{-g} F(R) + \int d^{4}x \sqrt{-g} \mathcal{L}_\mathrm{Matter} [g^{\mu \nu}, \Psi]
\label{action1}
\, ,
\end{align}
where $F(R)$ is a function of the Ricci scalar $R$ and $\kappa^{2} = 8\pi G = 1/M^{2}_\mathrm{pl}$.
$M_\mathrm{pl}$ is the reduced Planck mass, and $M_\mathrm{pl} \simeq 2 \times 10^{18}~[\mathrm{GeV}^{2}]$.
$\mathcal{L}_\mathrm{Matter}$ denotes the Lagrangian for a matter field $\Psi$.
The variation of the action \eqref{action1} with respect to the metric $g_{\mu \nu}$ leads to the equations of motion as follows,
\begin{align}
\label{jordaneom1}
F_{R}(R) R_{\mu \nu} - \frac{1}{2} F(R) g_{\mu \nu} + (g_{\mu \nu} \Box - \nabla_{\mu} \nabla_{\nu}) F_{R}(R) 
= \kappa^{2} T_{\mu \nu} (g^{\mu \nu}, \Psi)
\, .
\end{align}
$F_{R}(R)$ expresses the derivative of $F(R)$ with respect to $R$, $F_{R}(R) = \partial_{R} F(R)$, 
and hereafter we use this convention.
The energy-momentum tensor $T_{\mu \nu}$ is defined by
\begin{align}
T_{\mu \nu}(g^{\mu \nu}, \Psi) 
= \frac{-2}{\sqrt{-g}} \frac{\delta \left(\sqrt{-g} \mathcal{L}_\mathrm{Matter} (g^{\mu \nu}, \Psi) \right)}{\delta g^{\mu \nu}}
\, .
\end{align}
Furthermore, taking the trace of Eq.~\eqref{jordaneom1}, we obtain 
\begin{align}
\label{jordaneom2}
\Box F_{R}(R) 
= \frac{1}{3} \left[ 2 F(R) - R F_{R}(R)  + \kappa^{2} T  \right]
\, ,
\end{align}
where we have used conventions for the Ricci scalar as $R=R^{\mu}_{\ \mu}$
and for the trace of the energy-momentum tensor as 
$T = T^{\mu}_{\ \mu}$. 
We have to note that the trace  of the equations of motion is dynamical
although it is  $R = - \kappa^{2} T$ in the general relativity.

The scalar field is defined by the  identification 
\begin{align}
\label{identification}
\Phi \equiv& F_{R}(R) \, .
\end{align}
Solving the above relation with respect to $\Phi$, 
one can express the Ricci curvature $R$ in terms of $\Phi$, $R=R(\Phi)$.
Then, we can regard the right-hand side of Eq.~\eqref{jordaneom2} as the contribution from the potential term,
\begin{align}
\label{scalaron_potential1}
\frac{\mathrm{d} V(\Phi)}{\mathrm{d} \Phi} \equiv& \frac{1}{3} \left[ 2 F(R(\Phi)) - R(\Phi) F_{R}(R(\Phi)) \right]
\, ,
\end{align}
where $V(\Phi)$ is the potential of the scalar field $\Phi$. 
For this notation, Eq.~\eqref{jordaneom2} can be rewritten as the Klein-Gordon type equation:
\begin{align}
\label{jordanscalaroneom1}
\Box \Phi = \frac{\mathrm{d} V(\Phi)}{\mathrm{d} \Phi} + \frac{1}{3} \kappa^{2} T
\, .
\end{align}
It is also convenient to define the effective potential $V_{\mathrm{eff}}$ with the trace of energy-momentum tensor included:
\begin{align}
\label{jordaneffpotential_minimum}
\frac{\mathrm{d} V_{\mathrm{eff}}}{\mathrm{d} \Phi} 
=& 
\frac{1}{3} \left[ 2 F(R) - R F_{R}(R)  + \kappa^{2} T  \right]
\, .
\end{align}

The stationary condition $\mathrm{d} V_{\mathrm{eff}} / \mathrm{d} \Phi = 0$
gives the solution $\Phi=\Phi_{\min}$ at the potential minimum.
Defining $R=R_{\min}$ satisfying $\Phi_{\min}= F_{R} (R_{\min} )$, 
one obtains the mass of the scalar field defined as the second derivative of the effective potential;
\begin{align}
\label{jordanscalaronmass1}
m^{2}_{\Phi} 
=&
\left. \frac{\mathrm{d}^{2} V_{\mathrm{eff}}(\Phi, T)}{\mathrm{d} \Phi^{2}} \right|_{\Phi = \Phi_{\min}}
\nonumber \\
=&
\frac{1}{3} \left[ \frac{ F_{R}(R_{\min})}{F_{RR}(R_{\min})} - R_{\min}  \right]
\, .
\end{align}
It is of great importance that 
the effective potential $V_{\mathrm{eff}}$ in Eq.~\eqref{jordaneffpotential_minimum} includes the trace of energy-momentum tensor $T$.
Since the potential minimum changes according to the matter distribution,
the mass of scalar field Eq.~\eqref{jordanscalaronmass1}  shows the dependence of the environment.
In the next subsection, we will discuss this point in detail.

It is also notable that if Eq.~\eqref{jordaneom1} has a solution $R_{\mu \nu} = \Lambda g_{\mu \nu}$, 
we obtain the following relation:
\begin{align}
\label{stationary_condition}
2 F(R) - R F_{R}(R) + \kappa^{2} T = 0
\, ,
\end{align}
which corresponds to the stationary condition or potential minimum of the effective potential in Eq.~\eqref{jordaneffpotential_minimum}.
In the case of the positive $\Lambda$, 
we have the de Sitter solution in the $F(R)$ gravity. 
Thus, the $F(R)$ gravity can be a straightforward solution to the dark energy problem naturally giving rise to the cosmological constant \cite{Capozziello:2002rd}.

\subsection{The scalar field and the chameleon mechanism}
\label{sec2C}

The additional scalar field in $F(R)$ gravity can be responsible for the dark energy, to realize the de Sitter space-time as one of the solutions.
In the usual sense, the mass of the scalar field should be of the order of the Hubble scale $m_{\Phi} \sim 10^{-33}~[\mathrm{eV}]$.
However, such a light scalar field is easily detected and excluded by the observation because the Compton wavelength is very large.
On the other hand, 
if the scalar field is massive enough for its propagation to be frozen out, where the Compton wavelength is very small,
the scalar field can avoid the observational constraints.
However, such a massive scalar field is irrelevant for cosmology and gives us the inconsequential effect for the dark energy problem.
The screening mechanism gives us a solution to solve the conflict between those two different requirements on different scales.

In the chameleon mechanism, the scalar field potential depends on the energy-momentum tensor of the other matter fields,
which results in that the mass of the scalar field changes according to the matter fields surrounding the scalar field itself.
In other words, the scalar degree of freedom depends on the environment.
For example, the energy density characterizes the non-relativistic perfect fluid, and thus, the mass of the scalar field is a function of the energy density.
If one chooses the potential so that the mass increases at the high-density region and decreases at the low-density region,
the chameleonic scalar field makes the theory to be relevant in two different regions: cosmological scale and smaller scale.

As an illustration, we see how the chameleon mechanism works in a concrete model of $F(R)$ gravity,
as follows:
\begin{align}
\label{starobinsky_action1}
F(R) = R - \beta R_{c} \left[ 1 - \left( 1 +  \frac{R^{2}}{R^{2}_{c}} \right)^{-n} \right] + \alpha R^{2}
\, .
\end{align}
The $R_{c}$ expresses a typical energy scale where the modification of gravitation is relevant, 
and one expects $ R_{c} \sim \Lambda \simeq 4 \times 10^{-84}~[\mathrm{GeV}^{2}]$
so that the modification is responsible for the dark energy at the low-energy scale.
$\beta$ and $n$ are positive parameters.
Considering the large-curvature limit, $R>R_{c}$, the $F(R)$ function approximates as follows:
\begin{align}
\label{starobinsky_action2}
F(R) \approx R - \beta R_{c} + \beta R_{c} \left( \frac{R_{c}}{R} \right)^{2n} + \alpha R^{2}
\, .
\end{align}
The first and second terms in Eq.~\eqref{starobinsky_action2} imply that
this model recovers the $\Lambda$CDM model with the cosmological constant $2\Lambda = \beta R_{c}$
although the third term shows the modification in the low-curvature region, which is suppressed in the large-curvature region.
Note that if we expect $R_{c} \sim \Lambda$, it naively suggests $\beta = \mathcal{O}(1)$.

We introduce $\alpha$ to display another energy scale and assume that the $R^{2}$ term modifies the gravitational theory at a high-energy scale.
The $F(R)$ gravity models for the dark energy generally suffer from the curvature singularity problem \cite{Frolov:2008uf}.
The $R^{2}$ correction cures this problem \cite{Dev:2008rx, Kobayashi:2008wc}, and as a by-product of the cure for the singularity problem, the mass of the scalar wave is upper bounded \cite{Thongkool:2009js, Katsuragawa:2017wge}.
Hereafter, we rewrite $\alpha$ as $\alpha = \mu/R_{c}$, where $\mu$ is the dimensionless parameter, for the convenience.

For this $F(R)$ gravity model, one can calculate the potential $V(\Phi)$ as in Eq.~\eqref{scalaron_potential1}
\begin{align}
\frac{\mathrm{d} V(\Phi)}{\mathrm{d} \Phi} 
=
\frac{1}{3} 
\left[ 
R - 2 \beta R_{c} + 2 \beta R_{c} \left( 1 +  \frac{R^2}{R^{2}_{c}} \right)^{-n} 
 + 2 n \beta R_{c} \frac{R^{2}}{R^{2}_{c}} \left( 1 +  \frac{R^2}{R^{2}_{c}} \right)^{-(n+1)}  
\right]
\, , 
\end{align}
and the relation between $\Phi$ and $R$ is given by Eq.~\eqref{identification},
\begin{align}
\Phi(R) = 
1 - 2 n \beta \frac{R}{R_{c}} \left( 1 +  \frac{R^2}{R^{2}_{c}} \right)^{-(n+1)} + 2 \mu \frac{R}{R_{c}}
\, .
\end{align}
In order to derive the potential $V(\Phi)$, 
we replace the derivative of the potential with respect to $\Phi$ as follows:
\begin{align}
\frac{\mathrm{d} V(\Phi)}{\mathrm{d} R}
=& F_{RR}(R) \frac{\mathrm{d} V(\Phi)}{\mathrm{d} \Phi}
\, . 
\end{align}

\begin{figure}[htbp]
\centering
\includegraphics[width=0.6\textwidth]{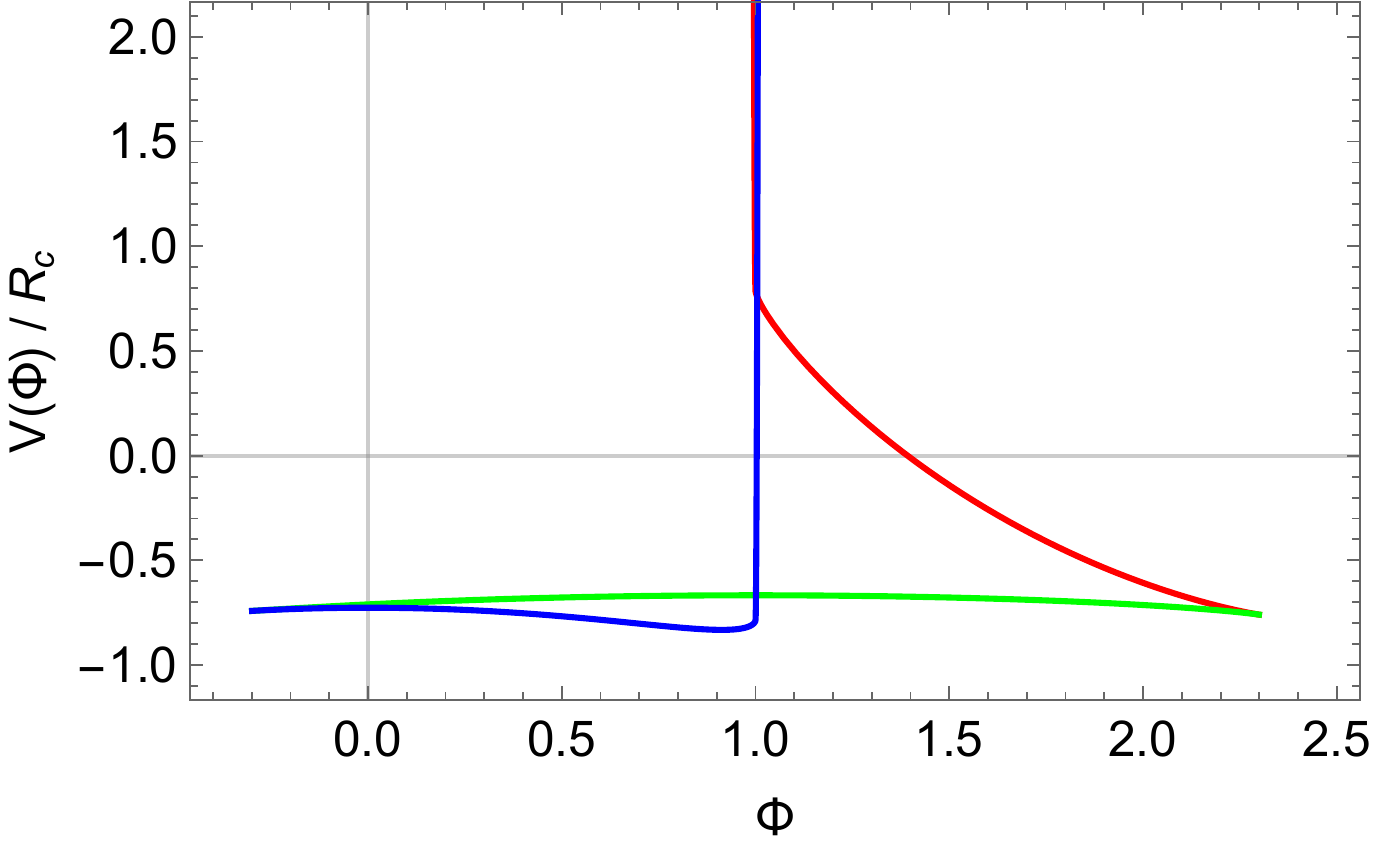}
\caption{
The potential of scalar field $V(\Phi)$ where $n=1$, $\beta=2$, and $\mu=10^{-6}$, normalized by $R_{c}$.
Potential branches at $R/R_{c} \approx \pm 1/\sqrt{3}$ where $F_{RR}(R)=0$. 
$\Phi=1$ on the green-solid line corresponds to an unstable flat space-time ($R=0$).
$\Phi=0$ on the blue-dashed line is the unstable de Sitter,
and the potential minimum on the blue-dashed line gives the stable de Sitter background.
}
\label{potential1}
\end{figure}

\begin{figure}[htbp]
\centering
\includegraphics[width=0.6\textwidth]{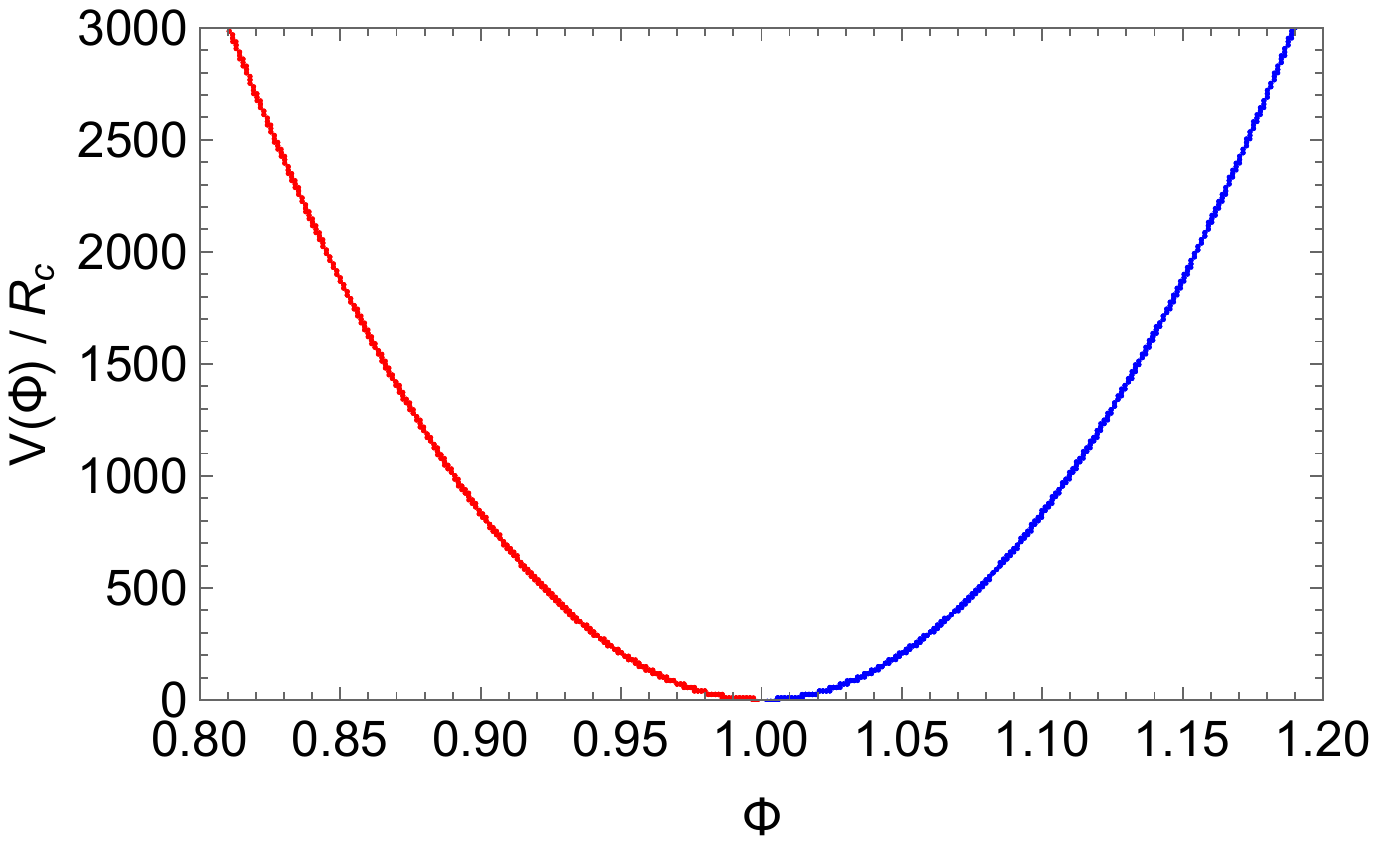}
\caption{
The same as in Fig.~\ref{potential1}.
When $R\rightarrow + \infty$, $\Phi \rightarrow + \infty$ (blue-dashed line).
When $R\rightarrow - \infty$, $\Phi \rightarrow - \infty$ (red-dotted line).
}
\label{potential2}
\end{figure}

As an example, we assume the model where $n=1$ and $\beta=2$ in Eq.~\eqref{starobinsky_action1}.
Integrating $\mathrm{d} V / \mathrm{d} R$ with respect to $R$, 
we obtain
\begin{align}
\frac{V(\Phi(R)) }{R_{c}}
=&
\frac{1}{6} 
\left[
16 \frac{R}{R_{c}} \left( 1 + \frac{R^2}{R^{2}_{c}} \right)^{-4}
- 40 \frac{R}{R_{c}} \left( 1 + \frac{R^2}{R^{2}_{c}} \right)^{-3}
\right. \nonumber \\
& \qquad 
+ \left(30 \frac{R}{R_{c}} + 8\right) \left( 1 + \frac{R^2}{R^{2}_{c}} \right)^{-2}
-\left(8 \mu \frac{R}{R_{c}} + 3 \frac{R}{R_{c}} + 12\right) \left( 1 + \frac{R^2}{R^{2}_{c}} \right)^{-1}
\nonumber \\
& \qquad \qquad \left. 
- 16 \mu \frac{R}{R_{c}} 
+ 2 \mu \frac{R^{2}}{R^{2}_{c}} 
+3 (8 \mu -1) \tan ^{-1} \left( \frac{R}{R_{c}} \right)
\right]
\end{align}
Note that the above expression is consistent with the result in \cite{Frolov:2008uf} without $R^2$ correction, 
up to the redefinition for the scalar field $\Phi$ and the integration constant for the effective potential $V_{\mathrm{eff}}$.
The parametric plot of $\left( \Phi(R) , V(\Phi(R)) \right)$ with respect to $R$
shows the form of the potential, given in Figs.~\ref{potential1} and \ref{potential2}.
The three branches (blue-dashed, green-solid, and red-dotted lines in Fig.~\ref{potential1}) show up
because $F_{R}(R)$, equivalently $\Phi$, is the multi-valued function with respect to $R$.

Next, we add the contribution from the matter field, that is, the trace of the energy-momentum tensor $T$ and consider the potential minimum and mass of the scalar field.
Because the chameleon mechanism works in the existence of the matter field, the corresponding space-time curvature is larger than the Hubble or dark energy scale.
Thus, we can utilize the large-curvature limit, $R>R_{c}$, in the following discussion.

In the large-curvature limit, the stationary condition in Eq.~(\ref{stationary_condition}) approximately leads to
\begin{align}
2 F(R) - R F_{R}(R) + \kappa^{2} T \approx
R - 2 \beta R_{c}
+ 2 (n + 1) \beta R_{c} \left( \frac{R_{c}}{R} \right)^{2n} + \kappa^{2} T
\, ,
\end{align}
and the curvature at the potential minimum is
\begin{align}
R_{\min} \approx - \kappa^{2} T 
\, .
\end{align}
And then, Eq.~(\ref{jordanscalaronmass1}) gives the mass of the scalar field at the potential minimum:
\begin{align}
\label{jordanmass2}
m^{2}_{\Phi} 
\approx \frac{1}{3} \left[ \frac{ 1 - 2 n \beta \left( \frac{R_{c}}{R_{\min}} \right)^{2n+1} + 2 \frac{\mu}{R_{c}} R_{\min} }
{\frac{2 n (2n + 1) \beta}{R_{c}} \left( \frac{R_{c}}{R_{\min}} \right)^{2n+2}  + 2 \frac{\mu}{R_{c}} } - R_{\min} \right]
\, .
\end{align}

When we use the pressure-less dust for the matter fields, the trace of the energy-momentum tensor is given as $T = - \rho$ where $\rho$ is the matter-energy density,
and the potential minimum is characterized by the curvature $R_{\min} = \kappa^{2} \rho $.
Because the mass \eqref{jordanmass2} is the increasing function of the curvature,
as we will see in Sec.~\ref{sec4C},
we can conclude that the scalar field becomes heavy in the high-density region.
Note that in the very large-curvature limit, or equivalently the very high-density region, 
the mass squared goes to $R_{c}/6\mu = 1/6\alpha$.


\section{Gravitational waves in $F(R)$ Gravity}
\label{sec3}

\subsection{The chameleon mechanism in scalar waves?}
\label{sec3A}

The additional scalar field in $F(R)$ gravity suggests a new polarization mode of the GWs
\cite{Capozziello:2008rq,Yang:2011cp,Rizwana:2016qdq,Gong:2017bru,Liang:2017ahj,Bogdanos:2009tn}.
Although one can describe the ordinary tensor modes of the GWs as the fluctuation around the background space-time,
the scalar wave is corresponding to the oscillation of the scalar field around the potential minimum
while the minimum is closely related to the background space-time as in Eq.~\eqref{stationary_condition}.
As we discussed in the previous section, the effective potential of the scalar field depends on the trace of the energy-momentum tensor, and thus, the oscillation around the potential minimum also changes according to the medium in which the GWs propagate.

Therefore, we naively expect that the chameleon mechanism affects the scalar mode of GWs.
In other words, we may find the environment dependence of the GW signal in $F(R)$ gravity.
Note that we study the perturbation at the first order and do not include the cross terms between the tensor and scalar modes as well as the back-reactions.


\subsection{Perturbations on curved space-time with matters}
\label{sec3B}

In order to discuss the chameleon mechanism in the scalar mode of GWs,
we first derive the basic equations for the GWs.
After the identification Eq.~\eqref{identification}, one finds the equations of motion given by
\begin{align}
\label{jordantensoreom}
&\Phi R_{\mu \nu} - \frac{1}{2} F(R) g_{\mu \nu} + (g_{\mu \nu} \Box - \nabla_{\mu} \nabla_{\nu}) \Phi
= \kappa^{2} T_{\mu \nu} (g^{\mu \nu}, \Psi) 
\, , \\
\label{jordanscalareom}
&\Box \Phi = \frac{\mathrm{d} V_{\mathrm{eff}}(\Phi)}{\mathrm{d} \Phi} 
\, .
\end{align}
To derive the equations for the GWs, 
we define the perturbation for the metric $g_{\mu \nu}$ and scalar field $\Phi$: 
\begin{align}
\label{jordantensorpert}
g_{\mu \nu} =& b_{\mu \nu} + h_{\mu \nu}
\, , 
\\
\Phi =& \Phi_{\min} + \phi
\, ,
\label{jordanscalarpert}
\end{align}
where $b_{\mu \nu}$ is the background metric,
and $\Phi_{\min}$ satisfies $dV_{\mathrm{eff}}/d\Phi = 0$.
Note that the perturbation of the scalar field $\phi$ can be written in terms of the metric perturbation $h_{\mu \nu}$
because the scalar field is, by definition, related to the space-time curvature as in Eq.~\eqref{identification}.

For the metric perturbations, 
the curvature $R$ and its function $F(R)$ can be expanded, up to the first order, around the background space-time as follows:
\begin{align}
R_{\mu \nu} &= R^{(b)}_{\mu \nu} + \delta R_{\mu \nu}
\label{jordanriemannpert} 
\, , \\
R &= R^{(b)} + \delta R 
\label{jordanriccipert}
\, , \\
\delta R &= b^{\mu \nu} \delta R_{\mu \nu} - h^{\mu \nu} R^{(b)}_{\mu \nu}
\, , \\
F_{R}(R) &= F_{R}(R^{(b)}) + F_{RR}(R^{(b)}) \delta R 
\label{jordanf(R)derivativepert}
\, .
\end{align}
Equations~\eqref{identification} and \eqref{jordanscalarpert} provide us with the following relations:
\begin{align}
&F_{R} (R^{(b)} ) = \Phi_{\min}\, , 
\\
&F_{RR} (R^{(b)} ) \delta R = \phi
\, .
\end{align}
According to the above relation, $F(R)$ is expanded around the minimum as follows:
\begin{align}
F(R) 
=& F(R^{(b)}) + \Phi_{\min} \delta R
\label{jordanf(R)pert}
\, .
\end{align}
And also, one finds that the Christoffel symbols are expanded by
\begin{align}
\Gamma^{\lambda}_{\ \mu \nu}
&=
\Gamma^{\lambda(b)}_{\ \mu \nu}
+ \frac{1}{2} b^{\lambda \rho} 
\left( \nabla^{(b)}_{\mu}h_{\nu \rho} + \nabla^{(b)}_{\nu}h_{\mu \rho} - \nabla^{(b)}_{\rho}h_{\mu \nu} \right) 
\, .
\end{align}

Finally, we obtain the perturbed equation of Eq.~\eqref{jordantensoreom} at the first order of perturbation:
\begin{align}
\label{tensormode1}
&\left( \nabla^{(b)}_{\mu} \nabla^{(b)}_{\nu} - b_{\mu \nu} \Box^{(b)} - R^{(b)}_{\mu \nu} \right) \phi 
\nonumber \\
& \qquad = 
\Phi_{\min} \delta R_{\mu \nu} 
- \frac{1}{2} \Phi_{\min} \delta R b_{\mu \nu} 
- \frac{1}{2} F(R^{(b)}) h_{\mu \nu} 
\nonumber \\
& \qquad \qquad
- \frac{1}{2} b_{\mu \nu}\left( 2 \nabla^{(b)}_{\alpha}h^{\alpha}_{\ \rho} - \nabla^{(b)}_{\rho}h \right) \nabla^{\rho}_{(b)} \Phi_{\min} 
- b_{\mu \nu} h^{\alpha \beta}  \nabla^{(b)}_{\alpha} \nabla^{(b)}_{\beta} \Phi_{\min} 
\nonumber \\
& \qquad  \qquad  \qquad
+ \frac{1}{2} \left( \nabla^{(b)}_{\mu}h_{\nu \rho} + \nabla^{(b)}_{\nu}h_{\mu \rho} - \nabla^{(b)}_{\rho}h_{\mu \nu} \right) 
\nabla^{\rho}_{(b)} \Phi_{\min}  
\, ,
\end{align}
where the covariant-derivative operators are composed by the background metric.
Note that we ignore the perturbation in the matter sector induced by the metric perturbation, $\delta T_{\mu \nu} = 0$,
where we treat the matter field as a fixed external field.
For the scalar field, we obtain the scalar wave equation
by taking the trace of Eq.~\eqref{tensormode1} or substituting the perturbations into Eq.~\eqref{jordanscalareom}
with the background equation:
\begin{align}
\label{scalarmode1}
\left( \Box^{(b)} - m_{\Phi}^{2} \right) \phi 
&= 
- \frac{1}{3} \Phi_{\min} h^{\mu \nu} R^{(b)}_{\mu \nu} 
+ \frac{1}{6} F(R^{(b)}) h 
\nonumber \\
& \qquad
+ \frac{1}{2} \left( 2 \nabla^{(b)}_{\mu}h^{\mu}_{\ \rho} - \nabla^{(b)}_{\rho}h \right) \nabla^{\rho}_{(b)} \Phi_{\min}
+ \frac{4}{3} h^{\mu \nu} \nabla^{(b)}_{\mu} \nabla^{(b)}_{\nu} \Phi_{\min}
\, .
\end{align}

It should be remarkable that
the effective potential is identical to the original potential $V_{\mathrm{eff}}(\Phi) = V(\Phi)$
if the energy-momentum tensor vanishes $T_{\mu \nu} = 0$.
Therefore, it is essential to assume the non-vacuum case;
otherwise, we cannot estimate the effect the chameleon mechanism on the scalar mode of the GWs.

\subsection{ Maximally symmetric background}
\label{sec3C}

Although we derived the basic equations for the GWs in $F(R)$ gravity in the curved space-time,
we assume that the background is maximally symmetric for simplicity,
that is, the Minkowski and (anti--)de Sitter background.
Based on this assumption and Eq.~\eqref{identification}, 
the background of the scalar field $\Phi_{\min}$ is constant
because the Ricci scalar $R$ is constant for the maximally symmetric space-time.
And then, the equations for the perturbation Eqs.~\eqref{tensormode1} and \eqref{scalarmode1} are simplified as
\begin{align}
\delta R_{\mu \nu} - \frac{1}{2} \delta R \, b_{\mu \nu} 
- \frac{1}{2} \frac{F(R^{(b)})}{\Phi_{\min}} h_{\mu \nu}
=& 
\left( \nabla^{(b)}_{\mu} \nabla^{(b)}_{\nu} - b_{\mu \nu} \Box^{(b)} - R^{(b)}_{\mu \nu} \right) \varphi \, , 
\label{tensormode2}
\\
\left( \Box - m^{2}_{\Phi} \right) \varphi 
=& 
- \frac{1}{3} \Phi_{\min} h^{\mu \nu} R^{(b)}_{\mu \nu} 
+ \frac{1}{6} F(R^{(b)}) h  \, ,
\label{scalarmode2}
\end{align}
where we define the new scalar field perturbation $\varphi \equiv \phi/\Phi_{\min}$, to normalize the original perturbation $\phi$.

For the comparison with the results in the previous works, we consider the Minkowski background.
In the flat space-time, the Ricci scalar vanishes, and $F(R)$ also vanishes in the present model Eq.~\eqref{starobinsky_action1}.
As a consequence, we obtain the same results in the literature (for example, see \cite{Capozziello:2008rq, Bogdanos:2009tn})
where GWs of  $F(R)$ gravity are dealt within the Minkowski background without matters.
We note, however, that the scalar field becomes tachyonic if we merely substitute $R=0$ into Eq.~\eqref{jordanscalaronmass1}
because the second derivative of $F(R)$ becomes negative.
Thus, the perturbation around the exact Minkowski background is unstable in our model of $F(R)$ gravity.
We may observe the similar behavior in other models of $F(R)$ gravity for dark energy \cite{Frolov:2008uf}.

On the other hand, we can improve the above discussion in other backgrounds.
Considering the stable de Sitter background corresponding to the potential minimum of the scalar field (see Fig.~\ref{potential1}),
we can address the dark energy problem at the cosmological scale.
If we choose this background, we can avoid the tachyonic scalar field and the consequent instability in the present model.
Furthermore, at the smaller scale,
we can safely ignore the effects on the perturbations induced by the cosmological background
and take $F(R^{(b)}) \approx 0$ and $R^{(b)}_{\mu \nu} \approx 0$ in the perturbed equations.
Thus, we can justify the previous analyses of the scalar wave around the Minkowski space-time within the approximation.

We also emphasize that the non-zero background curvature is unavoidable with
 respect to the stationary condition Eq.~\eqref{stationary_condition} 
in the presence of matter for the chameleon mechanism.
However, we can assume that the effect on the background space-time induced by the matter field is negligible
although we take it into account for the scalar field mass.
In the following analysis, 
we evaluate the mass of the scalar wave with non-zero space-time curvature to use the stable de Sitter minimum,
while we treat the background space-time approximately as the Minkowski space-time in the equations of perturbation.


\section{The Environment Dependence of Scalar Wave}
\label{sec4}

\subsection{Environment of ground-based GW observatory}
\label{sec4A}

It is difficult to simulate full processes of the screening mechanism in the scalar waves numerically because of the gigantic difference of the order of magnitude between the physical parameters; that is, the difference between the matter-energy density and dark energy density.
Instead of the exact situation, 
we use the toy model with several approximations and simplifications in order to evaluate the variation of the amplitude due to the chameleon mechanism in an analytical manner.

\begin{figure}[htbp]
\centering
\includegraphics[width=0.6\textwidth]{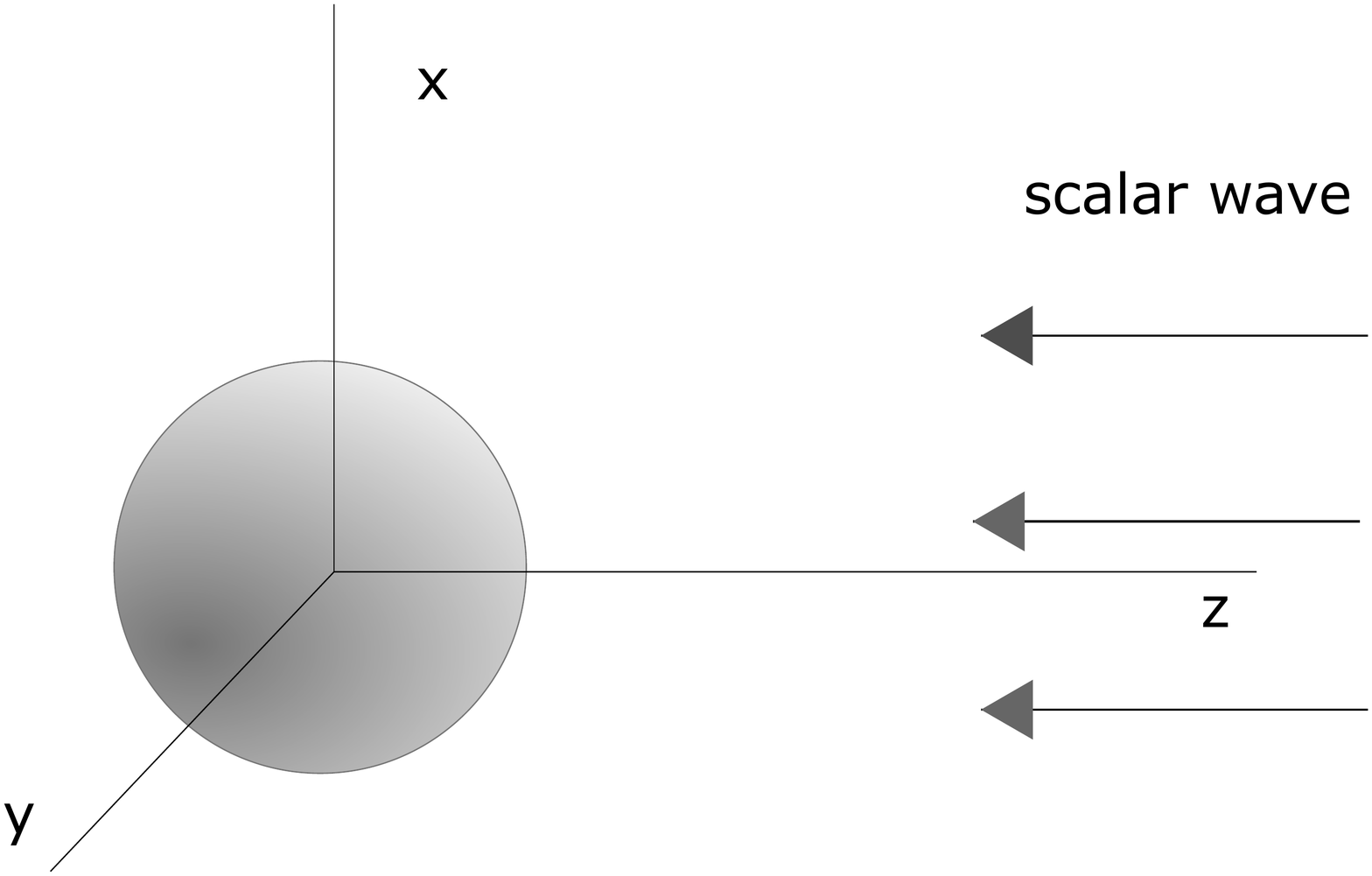}
\caption{The situation we consider now.
The plane scalar wave propagates to a spherically dense region from the positive direction of the $z$ axis.
}
\label{situation}
\end{figure}

We consider the following situation: 
the plane scalar wave is propagating from an infinitely far region to a spherically high-density region, as shown in Fig.~\ref{situation}.
We also assume the ground-based GW observatory locates inside the spherical region on the $z$ axis,
technically, near the boundary in the realistic situation.
Note that this model displays the simplified environment to reproduce our detection experiments: 
the scalar waves emitted far away from Earth propagate and enter the atmosphere, to be detected with the LIGO and Virgo gravitational observatories on Earth.
To simplify the analysis, we consider the following matter density profile:
\begin{equation}
\rho(x)=\rho_{a}-(\rho_a-\rho_{\infty})\Theta(r-r_{b})
\, ,
\end{equation}
where $\Theta(r)$ is the Heaviside step function.
$\rho_{a}$ and $\rho_{\infty}$ are the energy density of matters, satisfying the condition $\rho_a>\rho_{\infty}$.
$r$ is a radial coordinate, and $r_{b}$ denotes the radius of the high-density region.
For this situation, we obtain the following background solution:
\begin{equation}
\Phi_{bg}(r)=\left\{
\begin{array}{cc}
\Phi_{a}     &(0\leq r\leq r_{b})  \\
\Phi_{\infty}-(\Phi_{\infty}-\Phi_a) r_{b} \frac{e^{-m_{\infty}(r - r_{b})}}{r}     &(r_{b} < r) 
\end{array}
\right. ,
\label{phi_background}
\end{equation}
where $\Phi_i (i=\infty, a)$ denotes the minimum of the effective potential with respect to the energy density $\rho_i$.

\subsection{The suppression of scalar waves}
\label{sec4B}

For more simplification to address our toy model by hand,
we consider the region close to the $z$ axis,
where the scalar wave enters the high-density region almost vertically.
For this assumption, we can ignore the curvature of the spherical high-density region, and the boundary between the high-density and low-density regions looks flat.
Thus, we can approximate that the penetrating wave inside the high-density region is the plane wave as well as the incident wave.
In order to describe the above condition,
it is convenient to use the cylindrical coordinate
since the system of our interest has the axisymmetry.
Let $R$ represent the radial distance from the $z$ axis, $r^{2} = R^{2} + z^{2}$,
and the region close to the $z$ axis corresponds to the condition $R \ll r_{b}$,
which leads to the following expansion of the background $\Phi_{bg}$:
\begin{equation}
\Phi_{bg}(r)=\Phi_{bg}(\sqrt{z^2+R^2}) \simeq \Phi_{bg}(z) + \mathcal{O}\left(\frac{R}{r_{b}}\right)
\, .
\end{equation}
Thus, we can eliminate the $R$ dependence of the background field
if we consider the scalar mode of GWs penetrating the high-density region along to $z$ axis.
Provided the position of the boundary between the high-density region and space is $z=z_{b}(=r_{b})$,
$\Phi_{bg}$ in Eq.~\eqref{phi_background} can be approximated as follows: 
\begin{equation}
\Phi_{bg}(z)
= \left\{
\begin{array}{cl}	
\Phi_a&(|z| \leq z_b)\\
\Phi_{\infty}-(\Phi_{\infty}-\Phi_a)z_b\frac{e^{-m_{\infty}(z-z_b)}}{z}&(z_b < |z|)\\
\end{array}
\right.
\label{phi_background2}
\, .
\end{equation}

The equation of motion for the scalar wave in the cylindrical coordinate is given by
\begin{equation}
\left(-\frac{\partial^2}{\partial t^2} + \frac{\partial^2}{\partial z^2} \right) \phi(t, z)
= m^2_{\Phi}(z)\phi(t, z),
\label{full_scalar_eq}
\end{equation}
where $\phi(z)$ represents the plane scalar wave, and we ignore its $\rho$ dependence by the above condition.
In order to divide Eq.~\eqref{full_scalar_eq} into the equation of each frequency mode,
we consider the Fourier expansion of the plane wave $\phi (t, z)$,
\begin{equation}
\phi (t, z) = \int \frac{\mathrm{d}\omega}{2\pi}~\tilde{\phi}(\omega, z)~e^{-i \omega t}
\, ,
\end{equation}
and we obtain the equation of motion for the scalar wave in the simple form as
\begin{equation}
\frac{\mathrm{d}^2\tilde{\phi}}{\mathrm{d}z^2}
= \left(m^2_\Phi(\Phi_{bg}) - \omega^2\right)\tilde{\phi}
\, .
\label{scalar_eq_f}
\end{equation}
If we rewrite the right-hand side of the equation with $M(z)\equiv m^2_\Phi(\Phi_{bg})$, 
Eq.~\eqref{scalar_eq_f} takes the well-known form of the wave equation with the potential $M(z)$ in the (1+1) dimension. 

Finally, we can make a rough estimation for an amplitude of the scalar wave in the high-density region.
For scalar waves with the frequency $\omega < m_{\Phi}$,
the solution of Eq.~\eqref{scalar_eq_f} inside of the high-density region becomes
\begin{align}
\tilde{\phi}=C~e^{\sqrt{m^2_{\Phi}-\omega^2} z},
\label{amp_in_atmos}
\end{align}
where $C$ is a constant which represents the amplitude of the scalar wave,
and we dropped the growing mode of the solution in order to satisfy the boundary condition at the center.
One can find that the scalar waves of the low frequency, $\omega < m_{\Phi}$, is suppressed by the exponential factor.
We can see this property in a group velocity of scalar waves:
\begin{align}
v(\omega) = \frac{\sqrt{\omega^{2} - m_{\Phi}^{2}} }{\omega}
\, .
\label{group_velocity}
\end{align}
The scalar wave is decaying if $\omega < m_{\Phi}$,
which shows the behavior of our expectation due to the chameleon mechanism. 
In the following, we focus on this damping behavior inside the high-density region
although we will discuss the treatment of the boundary condition between the high- and low-density regions in conclusion.
It is notable that the chameleon mechanism does not affect the amplitude of the scalar mode of the high frequency $ \omega > m_{\Phi}$; however, it changes the phase for each Fourier mode.

\subsection{Estimation of scalar wave amplitude}
\label{sec4C}

\begin{figure}[htbp]
\centering
\includegraphics[width=0.6\textwidth]{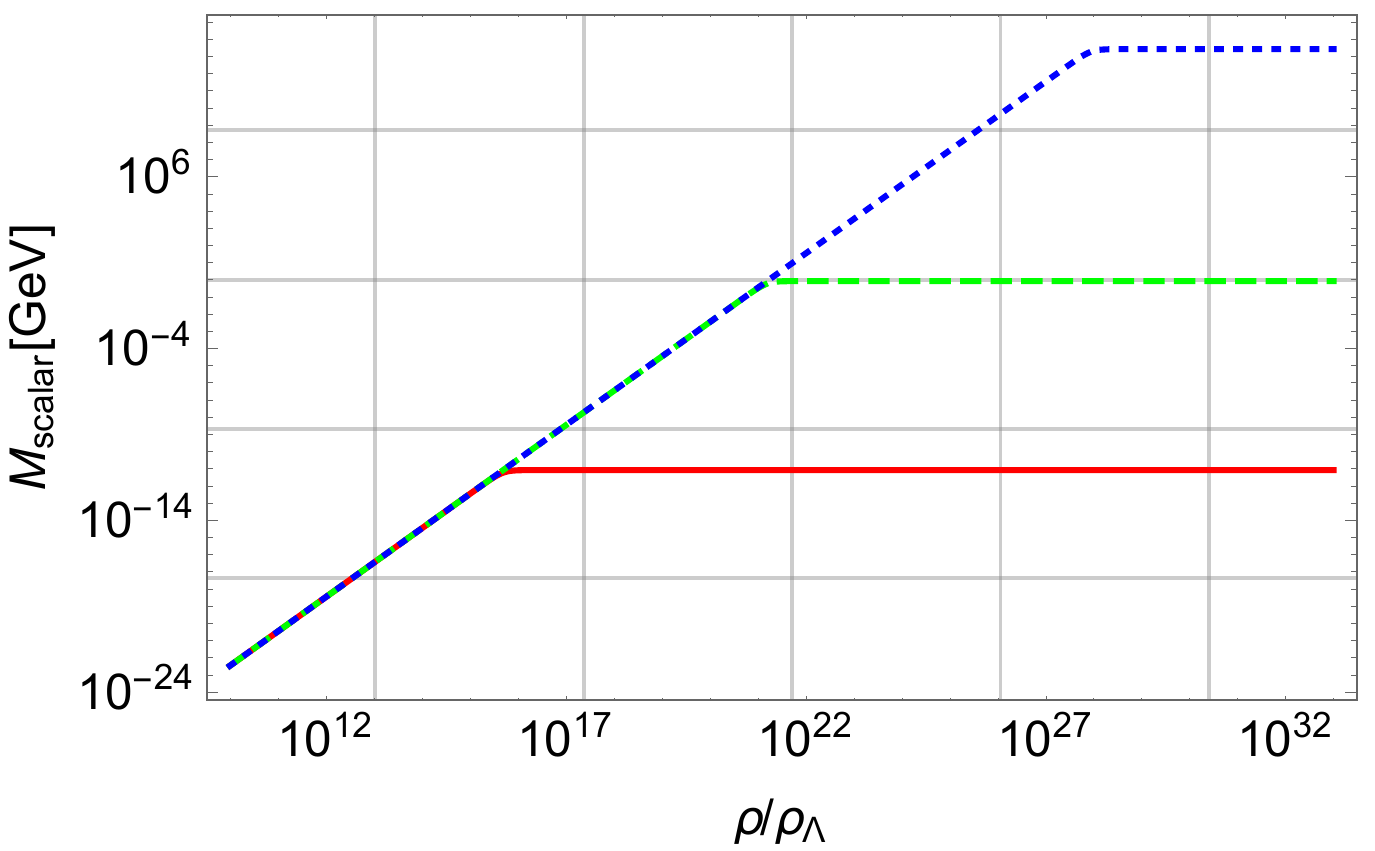}
\caption{
Mass of the scalar field with respect to the energy density $\rho$.
The energy density $\rho$ is normalized by the dark energy density 
$\rho_{\Lambda} \sim 6 \times 10^{-30}~\mathrm{[g/cm^{3}]}$.
Red-solid, green-dashed, and blue-dotted lines correspond to the choice of the parameter $\mu=10^{-62}$, $10^{-84}$, and $10^{-111}$, respectively.
Experimental bound is given by $m=\sqrt{\mu/6R_{c}} \geq 2.7 \times 10^{-12}~[\mathrm{GeV}]$ in \cite{Cembranos:2008gj},
which leads to $\mu \lesssim 10^{-62}$.
$\mu=10^{-84}$ corresponds to the mass $m \sim 1~[\mathrm{GeV}^{2}]$.
$\mu=10^{-111}$ corresponds to the inflaton mass $m \sim 10^{13}~[\mathrm{GeV}^{2}]$.
The other parameters are fixed as $n=1$ and $\beta=2$.
}
\label{mass_density}
\end{figure}

We further consider the damping factor in Eq.~\eqref{amp_in_atmos} in the actual case
where the high-density region is corresponding to the atmosphere of Earth.
We first calculate the mass of the scalar wave inside the atmosphere $m_{a}$ by Eq.~\eqref{jordanmass2}.
As an illustration, choosing the parameters as $\beta=2$, $n=1$ and $\mu=10^{-62}$,
the mass of the scalar wave in the atmosphere, $m_{\Phi} = m_{a}$, is given by
\begin{align}
m_{a} \simeq 8 \times 10^{-12}~\mathrm{[GeV]}
\label{criterion_mass}
\, ,
\end{align}
where we choose $\mu$ to respect the experimental upper bound \cite{Kapner:2006si,Adelberger:2006dh,Cembranos:2008gj}.
We also assume the energy density in the atmosphere $\rho = \rho_{a} = 10^{-9}~\mathrm{[g/cm^{3}}]$ 
at around $10^{5}~[\mathrm{m}]$ of the altitude.
We also plot the mass as the function of energy density for the particular choices of parameter $\mu$ in Fig.~\ref{mass_density}.
One can see that the mass becomes saturated in the high-density region and proportional to $\sqrt{1/\alpha} = \sqrt{R_{c}/\mu}$.

\begin{figure}[htbp]
\centering
\includegraphics[width=0.6\textwidth]{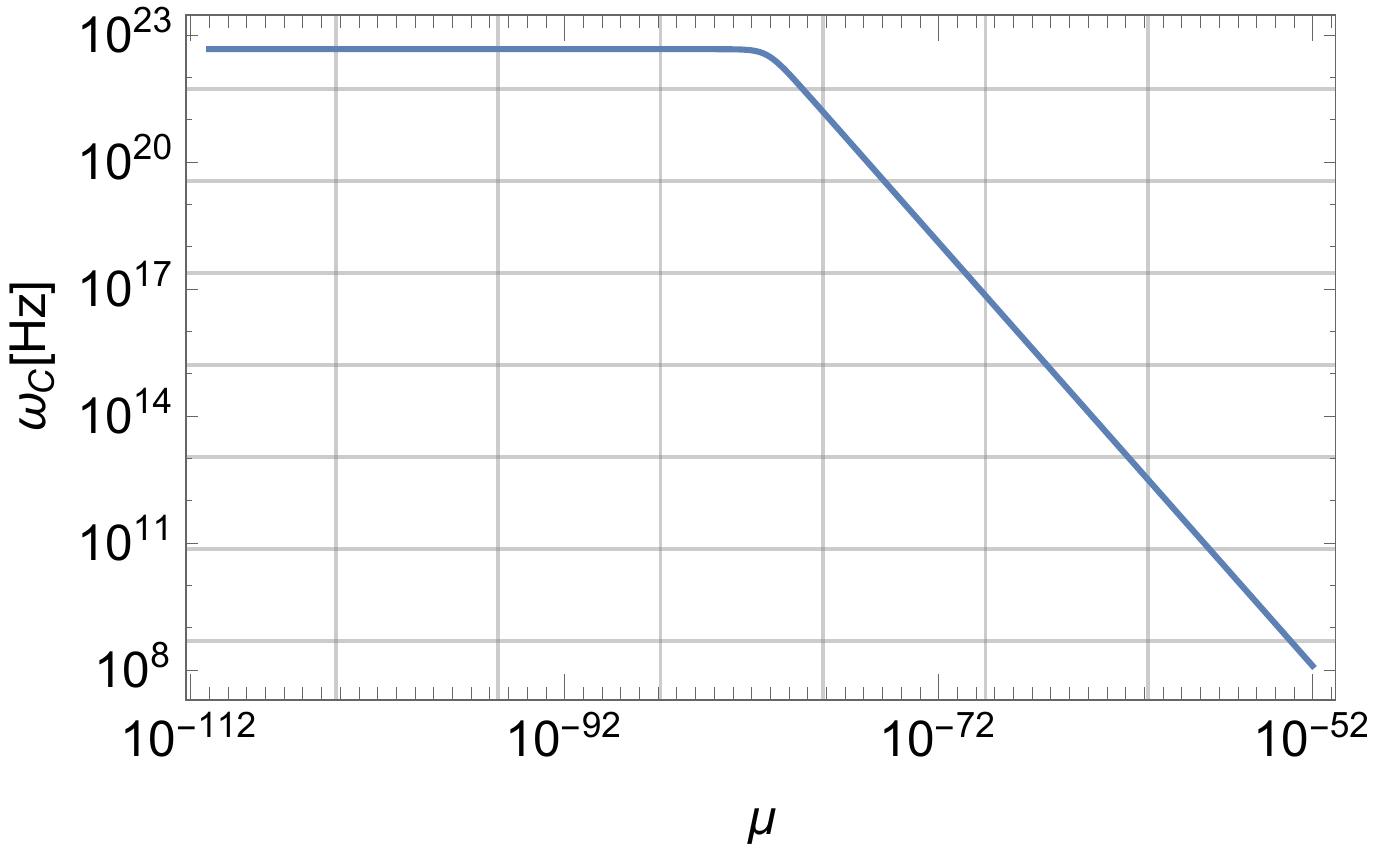}
\caption{
The criterion frequency with respect to the parameter $\mu$.
In contrast to Fig.~\ref{mass_density}, the energy density is fixed as 
$\rho = 10^{-9}~\mathrm{[g/cm^{3}}]$.
}
\label{omega_mu}
\end{figure}

Next, we compute the frequency that corresponds to $m_{a}$.
After the unit conversion, the frequency $\omega = \omega_{c}$ corresponding to the mass $m_{a}$ is given by 
\begin{align}
\omega_{c} \simeq 1 \times 10^{13}~\mathrm{[Hz]}
\label{criterion_frequency}
\, .
\end{align}
We call $\omega_{c}$ {\it criterion frequency }
because this value characterizes the two different behaviors of the scalar waves in different frequency regions
as we have seen in the previous subsection: 
the amplitude of the scalar wave decreases if the frequency is smaller than $\omega_{c}$,
while it is unchanged if the frequency is larger than $\omega_{c}$.
We also plot the criterion frequency as the function of parameter $\mu$ in Fig.~\ref{omega_mu}.
Note that Eq.~\eqref{criterion_frequency} gives us the lower bound of the criterion frequency
because Eq.~\eqref{criterion_frequency} corresponds the upper bound of $\mu$, 
and the criterion frequency becomes larger if we input smaller $\mu$. 

Even though the scalar modes with higher frequency than the criterion frequency, $\omega > \omega_{c}$, can avoid the gigantic suppression due to the chameleon mechanism, 
the criterion frequency is extremely high as in Eq.~\eqref{criterion_frequency}, and the detection is out of accessibility with our current technology.
Therefore, we can conclude that the detectable scalar waves should receive the sizable suppression before they reach the ground as far as our approximation is valid.

Note that 
we have ignored $\rho_{\infty}$ because we focus on the decaying scalar wave inside the atmosphere.
However, we can speculate that the scalar modes are also affected  by the chameleon mechanism during the propagation outside the atmosphere,
in analogy to the suppression inside the atmosphere,
because the energy density in space $\rho_{\infty}$ is finite.
As an example, we consider a simple case that $\rho_{\infty} = \rho_{\Lambda}$ 
to interpret the low-density region as the cosmological environment.
Because our model of $F(R)$ gravity is designed to explain the accelerated expansion of the Universe at the cosmological scale,
one can expect that the mass of scalar wave is of the order of the Hubble scale,
$m_{\Phi} \sim 10^{-33}~[\mathrm{eV}]$,  as we have discussed in Sec.~\ref{sec2C}.
Therefore, the criterion frequency is also of the order of the Hubble scale, 
which implies that the scalar waves emitted at astrophysical scales do not decay 
when they propagate in the cosmological environment.

Concerning the above estimation in the environment outside the atmosphere,
we mention that Ref.~\cite{Ip:2018nhl} performed a similar study on the high-frequency scalar modes in the context of the scalar-tensor theory and evaluated the amplitude of the scalar wave with spherical-wave expansions, which allows us to compare with our results.
Although the previous research discussed the suppression by the chameleon mechanism in the Milky Way halo,
we have considered the atmosphere of Earth, which is a much smaller scale than the galactic scale.
Furthermore, we have explicitly derived the damping factor as in Eq.~\eqref{amp_in_atmos} to show the exponential suppression in the amplitude for the scalar wave with the low frequency.

\subsection{Non-vertical waves}
\label{sec4D}

In the previous subsections, 
we have considered only the region $R \ll r_{b}$ and assumed that the incident wave enters the observatory vertically.
Under this assumption, we have found that the scalar waves with the detectable range of frequency receive strong suppression.
However, it is worth considering the non-vertically penetrating waves
because the interferometric GW detector does not have sensitivities to the vertical direction, along with the $z$ axis in our case,
for the scalar waves \cite{Maggiore:1999wm, Nakao:2000ug, Nishizawa:2009bf}.

If $R \gg r_{b}$, the penetrating waves are no longer plane, and we need to include spherical waves because we cannot neglect the curvature of Earth.
In this sense, it is straightforward that 
we consider the spherical-wave expansion of the scalar wave and solve Eq.~\eqref{scalarmode1} for each mode in the polar coordinate.
However, instead of the exact solution, 
intuitive calculations may allow us to estimate the contribution of the non-vertically penetrating waves.
In the following, we revisit the effect of the chameleon mechanism on the scalar waves by referring to the damping factor in Eq.~\eqref{amp_in_atmos}.

It is reasonable to assume that the damping factor in Eq.~\eqref{amp_in_atmos} appears even in the amplitude of the non-vertically penetrating waves
because the nature of the chameleon mechanism seems to have nothing to do with the kind of waves.
According to Eq.~\eqref{amp_in_atmos}, the damping factor is the increasing function of the distance, and thus, the scalar modes should show more screened behavior as they penetrate deeper into the atmosphere.
When the scalar modes penetrate the atmosphere and enter the observatory vertically, 
its trajectory is the shortest.
Thus, we estimate the value of the suppression factor when the scalar waves reach the ground vertically.

We assume that the frequency of the scalar mode $\omega$ is much smaller than the mass $m_{a}$ in the atmosphere;
for example, typical frequency $\mathcal{O}(100)~[\mathrm{Hz}]$ for the GWs from the binary mergers is smaller than the criterion frequency Eq.~\eqref{criterion_frequency}.
In this case, the frequency dependence in the damping factor is negligible in Eq.~\eqref{amp_in_atmos}.
Because the thickness of the atmosphere is $|z| \sim 10^5~[\mathrm{m}]$,
the damping factor is given as
\begin{align}
\exp[ - \sqrt{m^{2}_{a} - \omega^{2}} |z|]
\sim &
\exp[ - m_{a} |z|] 
\simeq
10^{- 2 \times 10^{9}}
\label{amp_in_atmos2}
\, .
\end{align}

Even in the shortest way, the scalar waves receive the gigantic suppression as in Eq.~\eqref{amp_in_atmos2} until they reach the GW observatory.
Because the non-vertically penetrating waves travel longer path as we have mentioned above, they are more suppressed, which implies further difficulties to detect non-vertical waves.

\subsection{Scalar waves from binary systems}
\label{sec4E}

In order to detect the scalar waves by ground-based GW observatories inside the atmosphere of Earth,
the scalar waves should have a sizable amplitude before they reach Earth
to overcome the extremely large suppression of Eq.~\eqref{amp_in_atmos2} by the chameleon mechanism.
Provided, however, that binary systems emit such scalar waves,
it may suggest the inconsistency with the existing constraints coming from the indirect detection of GWs.
In order to confirm the above statement, 
we roughly discuss the relation between the constraint from the binary and the amplitudes of the scalar mode of GWs.

First of all, let us comment on the formulation of the amplitude of GWs in  $F(R)$ gravity.
The luminosity formula in the context of $F(R)$ gravity is discussed in the  works 
\cite{DeLaurentis:2011tp, DeLaurentis:2013zv, DeLaurentis:2013fra}.
However, the formula therein is not straightforwardly applicable to the $F(R)$ gravity model in this paper
due to the issue of tachyon instability 
although, in  Ref.~\cite{DeLaurentis:2011tp},  the Minkowski background $R^{(b)} \simeq 0$ is considered.
It is worth noticing that the emission environment could accompany the nontrivial matter distribution, and thus, the exact Minkowski background is not suitable for the precise estimation.
Furthermore, we can expect that the amplitude of the GWs should depend on the emission environment, and the emission of the scalar waves would be suppressed due to the chameleon mechanism.
For example, the accretion disk surrounding the binary could make the scalar waves very massive and suppress the emission.
Although the energy fluxes with respect to the tensor and scalar modes are formulated in the context of the scalar-tensor theory in \cite{Alsing:2011er},
we need further assumptions to discuss the emission environment to respect the chameleon mechanism in terms of the mass of the scalar wave.

Irrespective of the emission environment for theoretical studies,
the observational data from the Hulse-Taylor binary gives the constraint on the emission of the extra mode of GWs \cite{Taylor:1982}.
The observational constraint on the orbital-period change is given in \cite{Will:2014},
\begin{align}
\frac{ \dot{P}_{\mathrm{corr}} }{ \dot{P}_{\mathrm{GR}}} = 0.997 \pm 0.002
\label{period_constraint}
\, , 
\end{align}
where $\dot{P}_{\mathrm{GR}}$ denotes the orbital-period change computed in the general relativity,
and $\dot{P}_{\mathrm{corr}}$ means the corrected one by subtracting the effect of rotation of the galaxy from the observed value.
When we assume that $\dot{P}_{\mathrm{corr}}$ includes that effects by the scalar field $\dot{P}_{\mathrm{scalar}}$
as $\dot{P}_{\mathrm{corr}} \rightarrow \dot{P}_{\mathrm{corr}} + \dot{P}_{\mathrm{scalar}}$ 
in Eq.~\eqref{period_constraint},
the effects on the orbital period should satisfy
\begin{align}
| \dot{P}_{\mathrm{scalar}} | \lesssim \mathcal{O}(0.001) | \dot P_{\mathrm{GR}} |
\label{scalar_period}
\, .
\end{align}
The above constraint suggests that
the deviation from the prediction in the general relativity should be small,
and thus, the emission of the additional polarizations of GWs, the scalar waves in our case, should also be suppressed.

Note that the frequency of the orbital period in the Hulse-Taylor binary is less than the criterion frequency $\omega_c$ 
as in Eq.~\eqref{criterion_frequency}.
Therefore, the amplitude of scalar waves is  suppressed by the factor in Eq.~\eqref{amp_in_atmos2} 
if the Hulse-Taylor binary emits scalar waves.
When we expect that the scalar-mode amplitude is of the same order of  tensor-mode amplitude at the detection, 
the former should be $10^{6 \times 10^{9}}$ times larger than the latter at the emission.
Because scalar waves are induced by  modified gravity,
such a large amplitude of the scalar wave implies a large deviation from the general relativity at the emission.
In the above case, the effects on the orbital-period change $\dot{P}_{\mathrm{scalar}}$ are no longer negligible, and it is against the observational constraint from the Hulse-Taylor binary.
Therefore, 
the large-scalar waves to overcome the gigantic suppression Eq.~\eqref{amp_in_atmos2} is tightly prohibited
in the case of the Hulse-Taylor binary.

Furthermore, if we assume 
that the above considerations for the Hulse-Tayor constraint are applicable to the other binaries which can be sources of the GWs detected by LIGO or Virgo,
the large emissions of scalar waves are also prohibited in any binaries,
and thus, the observed amplitude of the scalar mode is always much smaller than those of tensor modes in the ground-based detectors.  
Therefore, 
the current observational result that the scalar mode of GW has not been detected
is naturally understood 
because the chameleon mechanism suppresses the amplitude of the scalar wave in the atmosphere.
We emphasize that 
the chameleon mechanism hides any signals of the scalar waves in the current observational data, and it is almost impossible to find them from  binary mergers
unless there is a unique enhancement mechanism to overwhelm the suppression Eq.~\eqref{amp_in_atmos2}
during the propagation: that is after the emission from binaries and before the detection at ground-based GW observatories.


\section{Conclusions and Discussion}
\label{summary}

We have investigated the chameleon mechanism for the scalar mode of GWs in the framework of $F(R)$ gravity,
with the formulation of perturbations around a non-flat and maximally symmetric background.
We have modeled the particular environment to describe the GW detection by the observatory on Earth and to confirm the effect of the chameleon mechanism when the scalar waves penetrate the atmosphere of Earth.
Based on several approximations,
we have demonstrated that the scalar waves experience a huge damping factor during the propagation.
We have discussed the possibility to detect the scalar waves from the binary with the current ground-based GW observatories and explicitly examined our statement with the constraint on the scalar waves from the Hulse-Taylor binary.

As we discussed in Sec.~\ref{sec4}, 
the chameleon mechanism affects and suppresses the low-frequency modes.
From the theoretical viewpoint, we can regard the chameleon mechanism as a ``high-pass filter'' for the scalar waves.
Once we assume the spectrum of the initial scalar mode before entering the atmosphere, we can calculate the final spectrum by multiplying the filter with the initial spectrum.
This filter is the result of the chameleon mechanism, and thus, it depends on two elements: 
the effective potential of the scalar field and matter distribution in the atmosphere.
When we input the energy density of the atmosphere with respect to height,
we can eventually extract the information of the effective potential of the scalar field.
Moreover, we can generalize our analysis for a broader range of modified gravity theories, considering the chameleon mechanism.
For each theory, we can evaluate the characteristic frequency of the filter due to the chameleon mechanism.

From a phenomenological viewpoint,
the screening mechanism due to the dense atmosphere tightly constrains the detectability of the scalar mode.
Although we cannot observe the scalar mode in the single event with ground-based observatories,
the space-based observatories could provide us with another possibility to detect the scalar mode directly because we may avoid the screening mechanism in the space thanks to the detection environment with the low-energy density.
Then, we strongly propose the necessity of space-based observatories in view to detect such effects. 
Future space-based gravitational observatories give us potential detectability of the scalar mode,
to constrain the $F(R)$ gravity or other modified gravity theories with the chameleon mechanism.
If we can utilize observational data in which the chameleon mechanism is negligible, we can directly discuss the scalar mode of the GWs.

Furthermore, ground-based observatories also play a vital role 
when we can utilize the ground- and space-based observatories simultaneously.
We can imagine the following two cases:
(1) If ground-based observatories do not detect the scalar mode, but the space-based ones do,
it shows the environment dependence of the GW,
which suggests the chameleon mechanism.
(2) If both ground- and space-based observatories cannot detect the scalar mode,
there may be no signal of the physics beyond general relativity, and the $F(R)$ gravity theories are strongly constrained.
Of course, there is room for other possibilities:
the chameleon mechanism would already screen the scalar mode of GWs at the emission or propagation.

Besides the GWs from the binary, our analysis permits us to discuss the implication for the cosmology.
The non-tensorial modes of the GWs can be produced in the early Universe, 
which constitutes the stochastic background of the GWs at present.
The chameleon mechanism may also affect the scalar mode in this background at the detection, and accordingly, the search for the non-tensorial stochastic background of GWs is hard to confirm.
However, indirect detection is still possible, and we may observe the signal of the scalar mode embedded in the CMB.

In closing, let us give some comments on our future directions.
We have assumed several assumptions to address the chameleon mechanism.
We have utilized the step function to describe the matter distribution in the atmosphere
although the real distribution is continuous.
We may speculate that the effect of the chameleon mechanism is more powerful than that in the present analysis
because the density is higher as the altitude is lower in the atmosphere of Earth.
Therefore, the mass of the scalar wave is much larger than we have computed, which will be examined with the other prescriptions to reproduce the detection environment.

Moreover, we have not precisely addressed the boundary condition of the scalar wave at the atmosphere in the present paper.
We can discuss the junction condition of the scalar wave by evaluating that of the metric perturbations of space-time curvature,
according to the relation between the scalar field and the Ricci scalar.
However, since the length scale of the criterion frequency $\omega_{c}$ is much smaller than the thickness of the atmosphere,
we can expect that the amplitude is suppressed and decaying enough until the scalar waves reach the ground and that our kinematic analysis is roughly valid at the linear level.
The dynamical behavior of the scalar mode around the boundary at the nonlinear level
is still unknown
because it requires a full numerical analysis.
To conclude,
we have to observe the decaying scalar waves
due to the chameleon mechanism around the boundary.
Although realistic parameters for the environment can be defined,
the large difference of magnitude among those parameters could make the numerical analysis unmanageable,
which will be addressed in the framework of the other toy models.


\section*{Acknowledgments}
Authors are deeply indebted to Hiroyuki Nakano and Shin'ichi Nojiri for their helpful discussions and suggestions.
T.K. is supported by International Postdoctoral Exchange Fellowship Program at Central China Normal University and Project funded by China Postdoctoral Science Foundation 2018M632895, 
and is grateful to Taotao Qiu for the comments on the stochastic background.
T.I. acknowledges the financial support provided under the European Union's H2020 ERC Consolidator Grant 
``Matter and strong-field gravity: New frontiers in Einstein's theory'' Grant Agreement No. MaGRaTh-646597, 
and under the H2020-MSCA-RISE-2015 Grant No. StronGrHEP-690904. 
T.I. also thanks Christopher. J. Moore for useful discussion.
S.C. thanks the COST Action CANTATA (CA-15117) and INFN (iniziativa specifica QGSKY) for partial financial support.


\end{document}